\documentclass[eptcs]{eptcs}

\usepackage{breakurl}

\usepackage{amsmath}
\usepackage{amssymb}
\usepackage{graphicx}
\usepackage{subfigure}
\input xy
\xyoption{all}

\usepackage{meaning}
\usepackage{mathtools}
\usepackage{setspace}

\makeatletter
\newcommand{\figcaption}[1]{\def\@captype{figure}\caption{#1}}
\newcommand{\tblcaption}[1]{\def\@captype{table}\caption{#1}}
\makeatother

\title{RNA-interference and Register Machines \\
(extended abstract)}
\author{Masahiro Hamano
\institute{PRESTO
, Japan Science and Technology Agency (JST)
}
\institute{4-1-8 Honcho Kawaguchi, Saitama 332-0012, JAPAN.}
\email{hamano@is.s.u-tokyo.ac.jp}
}
\def\titlerunning{$\RNAI$ and Register Machines}
\def\authorrunning{Masahiro Hamano}

\newtheorem{thm}{Theorem}[section]

\newtheorem{prop}[thm]{Proposition}

\newtheorem{definition}[thm]{Definition}

\newcommand{\Inc}[1]{{\sf Inc}{(#1)}}
\newcommand{\RNAI}{{\sf RNAi}}
\newcommand{\RNARI}{{\sf recRNAi}}
\newcommand{\mRNA}{{\sf mRNA}}
\newcommand{\abmRNA}{{\sf mRNA}^{ab}}
\newcommand{\DNA}{{\sf DNA}}
\newcommand{\RNA}{{\sf RNA}}
\newcommand{\dsRNA}{{\sf dsRNA}}
\newcommand{\RISC}{{\sf RISC}}
\newcommand{\siRNA}{{\sf siRNA}}
\newcommand{\Dicer}{{\sf Dicer}}
\newcommand{\RMI}{{\bf RM}_\RNAI}
\newcommand{\RMRI}{{\bf RM}_{\sf recRNAi}}
\newcommand{\RdRp}{{\sf RdRp}}
\newcommand{\Dec}[2]{{\sf DecJump}(#1,#2)}

\begin{document}
\maketitle

\begin{abstract}
$\RNA$ interference ($\RNAI$) is a mechanism whereby small $\RNA$s
($\siRNA$s) directly control gene expression without
assistance from proteins. This mechanism consists of interactions between
$\RNA$s and small $\RNA$s both of which may be single or double
 stranded. The {\em target} of the mechanism is $\mRNA$ to
be degraded or aberrated, while the {\em initiator} is double
 stranded $\RNA$ ($\dsRNA$)
to be cleaved into $\siRNA$s.
Observing the digital nature of $\RNAI$, we represent $\RNAI$ as a
Minsky register machine
such that (i) The two registers hold single
and double stranded $\RNA$s respectively, and (ii) Machine's instructions are
interpreted by interactions of enzyme (Dicer), siRNA (with $\RISC$
 complex) and polymerization (RdRp) to the appropriate registers.
Interpreting
$\RNAI$ as a computational structure, we can investigate the computational meaning
of $\RNAI$, especially its complexity.
Initially, the machine is configured as a Chemical Ground Form
(CGF), which generates incorrect jumps.
To remedy this problem, the system is remodeled as recursive $\RNAI$,
in which $\siRNA$ targets not only $\mRNA$ but also
the machine instructional analogues of $\Dicer$ and $\RISC$.
Finally, probabilistic termination is investigated in the recursive $\RNAI$ system.
\end{abstract}

\section{Introduction}
\begin{figure}[!htb] 
\tiny
$$ \xymatrix@!0@C=0.6in@R=1.5pc
{ & &  & \mbox{\sf \small dsRNA} \\
& & 
\ar@{*{|}*{|}*{|}}[rr]  \ar@{-}[rr] <3pt> 
  \ar@{*{|}*{|}*{|}}[rr] <-10pt> \ar@{-}[rr] <-13pt>  & & 
  \ar@{=>}@/^1pc/[ddr]^{\mbox{\normalsize Dicer}}="D"
    &   \\ \\
\ar@<0pt> [r]^{\mbox{\footnotesize RdRp}}
& \ar@{=>}@/^1pc/[uur]  <2pt>  
 &   &  &  &
\ar@{*{|}*{|}*{|}}[l] <13pt> |\hole \ar@{-}[l] <16pt>  |\hole
 \ar@{*{|}*{|}*{|}}[l] <0pt> |\hole \ar@{-}[l] <-3pt>  |\hole 
& 
 \ar@{*{|}*{|}*{|}}[l] <13pt> 
|\hole \ar@{-}[l]<16pt>   |\hole
  \ar@{*{|}*{|}*{|}}[l] <0pt> 
|\hole \ar@{-}[l] <-3pt>_{\mbox{\small
siRNA's}}  |\hole 
  \\   
   \ar@{*{|}*{|}*{|}}[rr]<9pt> \ar@{-}[rr]<6pt>_{\mbox{\small $\abmRNA$}}  & 
  & & & &  
  \ar@{=>}@/^1pc/[ddl]^{
 \stackrel{\mbox{\footnotesize argonaute}
}{\frac{\mbox{\huge $\frown$}}
{||||||||||||}
}
\mbox{\small \sf RISC}}
 \\ 
\\
  &    
  &  \ar@{=>}@/^1pc/[uul]^{
   \mbox{\small aberration}}  
   &  & 
    \\  
 & &    
  \ar@{*{|}*{|}*{|}}[rr]<9pt>^{
\frac{
\mbox{\huge $\frown$}
}{|||||||||||||}
}
 \ar@{-}[rr] <6pt>  
  &   \mbox{\sf \small mRNA}  \ar@{=>} @/_1pc/ [dr]
  \ar@{<=} @/^1pc/ [dl]
  &
  \\
 & &
\mbox{\footnotesize transcription}
  & &  \mbox{\normalsize  $0$} & & 
}$$
\caption{RNA interference}
\label{circ}
\end{figure}
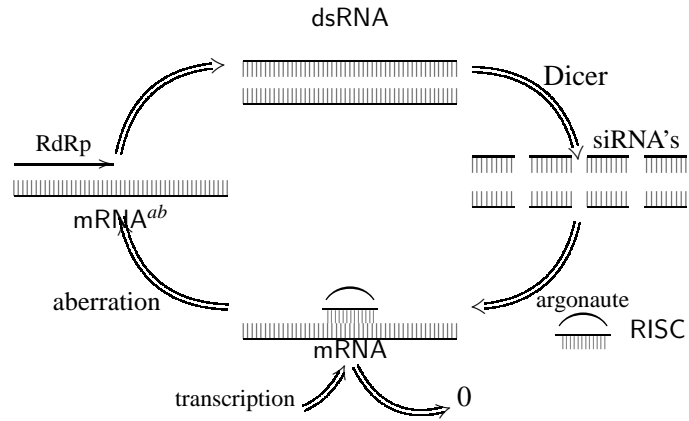

RNA interference ($\RNAI$), also known as RNA silencing,
is a mechanism whereby
a small interfering $\RNA$ ($\siRNA$) originating
from double stranded $\RNA$ ($\dsRNA$) directly controls gene expression of
a target $\mRNA$ \cite{BauN,Fire}.
The two key steps of $\RNAI$ are: \\
(i) $\dsRNA$ is cleaved into small
$\siRNA$'s fragments by an enzyme known as $\Dicer$. \\ (ii)
A single strand of one small $\siRNA$ is recruited by the {\sf argonaute}
protein to form a complex called $\RISC$.
Using the $\siRNA$ as a template,
$\RISC$ then identifies matching sequences in a target $\mRNA$, and induces the $\mRNA$
to degrade or become aberrant
(see the right semicircle of Figure \ref{circ}). \\
Therefore, we can regard the
initiator of $\RNAI$ as $\dsRNA$ (since it supplies the $\siRNA$s)
and the
target as $\mRNA$ (to be degraded or aberrated by a $\siRNA$
in a Watson-Crick complementary manner).

A third step of $\RNAI$ completes a circular pathway 
from the target to the initiator
\cite{BV,Jorgen}: \\ (iii) An aberrant $\mRNA$ resulting from step (ii) becomes a template for $\dsRNA$
produced by polymerization of RNA-dependent RNA polymerase ($\RdRp$)
(see the left semicircle of Figure \ref{circ}).

Since each step is digital and circularly linked, $\RNAI$ resembles a kind of (digital) computation.
This observation raises the question
of whether $\RNAI$ can be viewed as a digital computation.
If so, what is a computational meaning of
$\RNAI$ and how computationally complex $\RNAI$ is.
The purpose of this paper is to address these issues.

Firstly, we observe that $\RNAI$ can be modeled as a Minsky
register machine. The Minsky register machine is a Turing complete
model of computation, that (instead of an infinite tape for Turing
machine) is equipped with two registers (for holding numbers)
and a finite number of instructions (increment and decrement/jump)
acting on the registers \cite{Minsky}. While most biological
computational models to date are based on the Turing machine model; that is, they regard $\DNA$ as analogous to a single tape \cite{DNA},
the Minsky machine interpretation proposed here is intrinsic to the $\RNAI$ mechanism, whereby
$\RNA$s can be single or double
stranded. 
We first present a naive machine model of $\RNAI$, designated $\RMI$, 
in which the two registers are realized respectively as
the initiator ($\dsRNA$) and the target ($\mRNA$) of $\RNAI$.
Increment/Decrement instructions on the registers represent
chemical reactions mediated by enzymes and
proteins (e.g., $\RdRp$ and transcription/$\Dicer$ and $\RISC$).
However, the naive model lacks any rigorous computational language,
hence requires a syntactical analysis. 
Capturing $\RNAI$ as a computational structure, such analysis aims to
extract the computational meaning, in particular, the complexity, of $\RNAI$.

Motivated by the work of Zavattaro-Cardelli \cite{ZC},
we describe our machine $\RMI$
in the calculus of Chemical Ground Form (CGF), 
which is a minimal fragment of Milner's CCS equipped with interaction rates for
each channel, and hence constitutes a subset of the stochastic $\pi$-calculus \cite{Pri}.
Introduced by Cardelli \cite{Car}, 
CGF represents chemical kinetics
by giving correspondence to
a stochastic semantics of continuous
time Markov chains.
Despite its simplicity, the model sufficiently describes chemical
kinetics compositionally.
However, 
the primitive description of CGF lacks any direct representation
of zero-tests for the registers,
creating a tendency for the instructions
of encoded $\RMI$ to allow incorrect jumps.
To avoid such erroneous probabilistic jumping, an inhibitor
must be incorporated into the machine instructions.
Biologically, this corresponds to a process known as {\em recursive} $\RNAI$ ($\RNARI$),
an extension of $\RNAI$ 
\cite{recursive,predpl,nf}, whereby $\siRNA$ produced and accumulating
during $\RNAI$ inhibits not only $\mRNA$ but also $\RISC$ and $\Dicer$.
The extension to {\em recursive} $\RNAI$ ($\RNARI$) is obtained by adding a feedback linkage to
$\RNAI$. The $\RNARI$ is directly represented by a register machine $\RMRI$, 
in which $\siRNA$s interactions are naturally interpreted
as instruction inhibitors.
We describe the machine in terms of
CGF with fixed points.
Probabilistic termination is then investigated in the $\RNARI$ encoded system, and
Turing completeness up to any degree of precision is demonstrated.

\section{A Naive Interpretation 
of $\RNAI$
in Minsky Register Machine} \label{naive} 
In this section, we show that $\RNAI$ is naively
interpreted as Minsky register machine \cite{Minsky}. 
\begin{definition}[Register machine $\RMI$ interpreting $\RNAI$
 (cf. Figure \ref{MR})]
\label{rmi} {\em
$\RNAI$ is interpreted in the Minsky register machine $\RMI$
as follows:
Registers $r_1$ and $r_2$ hold species $\dsRNA$ and $\mRNA$ respectively
so that the increment on $r_1$ (res. $r_2$) produces one $\dsRNA$
(res. one $\mRNA$) and the decrement on $r_1$ (res. $r_2$) removes one $\dsRNA$
(res.one $\mRNA$). 
In biological terms, the increment on register $r_1$ represents
{\em polymerization} {\sf RdRp} with an aberrant $\mRNA$ template,
while an increment on $r_2$ represents {\em transcription}.
A decrement on $r_1$ models
the {\em enzyme} $\Dicer$ which cleaves $\dsRNA$ into $\siRNA$s,
and a decrement on $r_2$ models the complementary {\em
degradation} of $\mRNA$ by $\RISC$. \footnote{
The machine interpretation
assumes that
the two species of $\dsRNA$ and $\mRNA$ are disconnected, so that the
 decrement and increment of either species induces no effect on the other.
This assumption is justified because
the synthesis of $\dsRNA$ is here regarded as
primer-{\em independent} only \cite{BauN,BV}; in other words, $\dsRNA$
 is directly duplicated in the absence of primer.
In primer-{\em dependent} $\dsRNA$ synthesis,
the disconnection of the two species is violated. In such
scenario, $\siRNA$
triggers polymerization, hence enables $\RdRp$ to copy a normal $\mRNA$.
See also the author's \cite{HamSASB} on the difference of the two syntheses.
}
}\end{definition}

\begin{figure}[!htb]
$\resizebox{7.7cm}{!}{
\begin{tabular}{c|c|cc}  \hline
register   &   values   & increment/decrement   \\ \hline
$r_1$        & $\overbrace{{\sf dsRNA} \mid \cdots \mid {\sf dsRNA}}^{m_1}$  & 
${\sf Inc}(r_1):=RdRp$ 
&
\tiny \xymatrix{
\ar@{*{|}*{|}*{|}}[rr]  \ar@{-}[rr] <3pt> 
 \ar@<-10pt> [r]_{RdRp}     &  & 
}
 \\ & &  \\
& &  ${\sf Dec}(r_1):=Dicer$ &
\tiny \xymatrix@!0@C=0.5in@R=1.5pc
{ &
\ar@{*{|}*{|}*{|}}[l] <13pt> |\hole \ar@{-}[l] <16pt>  |\hole
 \ar@{*{|}*{|}*{|}}[l] <0pt> |\hole \ar@{-}[l] <-3pt>  |\hole &
\ar@{*{|}*{|}*{|}}[l] <13pt> |\hole \ar@{-}[l] <16pt>  |\hole
 \ar@{*{|}*{|}*{|}}[l] <0pt> |\hole \ar@{-}[l] <-3pt>  |\hole 
}
\\ & &  \\ \hline
\end{tabular}
}$
\hfill
$\resizebox{7.7cm}{!}{
\begin{tabular}{c|c|cc}  \hline
register   &   values   & increment/decrement   \\ \hline
$r_2$        & $\overbrace{{\sf mRNA} \mid \cdots \mid {\sf mRNA}}^{m_2}$  & 
${\sf Inc}(r_2):=\mbox{transcription}$  &  
\tiny \xymatrix@!0@C=0.6in@R=1.5pc {   \\ 
\ar@{*{|}*{|}*{|}}[rr] <20pt> \ar@{-}[rr] <17pt>   & & }
\\
                &  & ${\sf Dec}(r_2):=\RISC $  &
\tiny \xymatrix{   & \frac{\mbox{$\stackrel{\mbox{\Huge
 $\frown$}}{
}
$}}{||||||||}  \\
\ar@{*{|}*{|}*{|}}[rr] <20pt> \ar@{-}[rr] <17pt> 
 & 
&  
 }
\\  \hline
\end{tabular}
}$
\caption{Register Machine $\RMI$}
\label{MR}
\end{figure}
\noindent
The following table displays the chemical reactions for the
increment/decrement on the two registers, where
$\mRNA^{\bullet}$ denotes either $0$ or $\abmRNA$.

\begin{table}[htbp]
$\resizebox{15cm}{!}{
\begin{tabular}{c|lc|lc}
   & \multicolumn{2}{c}{$r_1$} \vline   & \multicolumn{2}{c}{$r_2$} \\ \hline
increment & 
(polymerization) & 
\mbox{\small $\RdRp + \abmRNA \longrightarrow \dsRNA$}
&  
{(transcription)}  &
 \mbox{ $\longrightarrow \mRNA$}
 \\ \hline
decrement  &    (cleavage) & 
\mbox{$\dsRNA + \Dicer \longrightarrow \siRNA 's$} &
(degradation)  &
\mbox{$\mRNA + \RISC \longrightarrow \mRNA^{\bullet} + \RISC$}
\end{tabular}
}$
\caption{chemical reactions}
\label{chemreac}
\end{table}

\section{$\RNAI$ as Chemical Reaction and Register Machines} \label{secchem}
In this section, we describe the register machine $\RMI$ in Section
\ref{naive} in terms of 
of CGF. 
Recall that CGF is a subset of $\pi$-calculus and of CCS 
supplemented with channel transition rates.
Using three interaction prefixes
$\pi := \tau_{(r)}$, $? a_{(r)}$ and $! a_{(r)}$,
CGF models collision between molecules as well as
molecular decay. 
The parenthesized subscript $(r)$ denotes the reaction rate of the
channel. Collision and decay are described by \\
$\begin{array}{lccc}
\mbox{(decay of molecule)} &
\cdots \oplus \tau_{(r)}. Q  \oplus \cdots
&
\longrightarrow
& 
Q \\
\mbox{(collision of molecules)} &
\cdots \oplus ?a_{(r)}. Q  \oplus \cdots \bigm |
 \cdots \oplus !a_{(r)}. R  \oplus \cdots 
&
\longrightarrow
& 
Q \bigm | R
\end{array}$ \\
Then a CGF is a pair $(E,P)$ of a set $E$
of {\em reagents} and a initial {\em solution} $P$.
A reagent $X_i = M_i$ for naming a chemical specie and {\em molecules}
$M_i$ for describing the interaction capabilities of the corresponding
species.
Solution is a multiset of variables, which is released  by interactions: \\
$
\begin{array}{ccc}
\mbox{(Reagents)~}  E :=  0 \mbox{~and~} X=M,E &
\mbox{~(Molecule)~}  
M  :=  0 \mbox{~and~} \pi. P \oplus M   &
\mbox{~(Solution)~}  
P :=   0 \mbox{~and~} X \bigm |  P 
\end{array}
$ \\
Formally, computation of CGF is defined in terms of Labelled Transition Graph,
as defined in \cite{Car}. 


\smallskip

Every increment instruction
$I_i=\Inc{r_j}$ is formalized directly for $j \in \{1,2 \}$
so that once the chemical reactions of the first row of Table
\ref{chemreac} are complete,
we proceed to the next instruction $I_{i+1}$.

\noindent (Increment $I_i=\Inc{r_j}$)
$$\begin{array}{ccc}
I_i = &  \RdRp \bigm| \tau. I_{i+1}  &  j=1  \\ 
I_i = &  \mRNA \bigm| \tau. I_{i+1} &   j=2
\end{array}$$

\smallskip
The decrement operations are more subtle.
Decrements on on $r_1$ and on $r_2$
represent the chemical reactions of the second row
of Table \ref{chemreac},
which reactions ensure that $\Dicer$ and $\RISC$
interact to the entities in $\dsRNA$ and $\mRNA$
respectively, and thereby eliminate them.
Although $\Dicer$ and $\RISC$ both induce
decremental operations,
$\RISC$ is recycled during degradation so that it is retained
in the right-hand-side of (degradation),
while the $\Dicer$ catalyst
is consumed during the reaction (cleavage).


So that the registers may be decremented, they are interpreted as
follows:
{
\setlength{\belowdisplayskip}{0pt}%
\setlength{\abovedisplayskip}{0pt}
$$\begin{array}{lcc}
\mbox{Register $r_1$} & 
\hspace{2cm} \dsRNA & := ? a_1. (\siRNA \bigm | \cdots \bigm | \siRNA) \\
\mbox{Register $r_2$} &
\hspace{2cm} \mRNA & :=  ? a_2 . (\tau. 0\oplus \tau. \abmRNA)
\end{array}$$
}
They represent  that $\dsRNA$ and $\mRNA$ disappear by
formation of $\siRNA$, and by degradation or aberration, respectively.

If the chemical reaction occurs in the presence of $\dsRNA$ (res. $\mRNA$),
we proceed to the instruction $I_{i+1}$.
Otherwise (i.e. if the reaction does not occur because
$\dsRNA$ is absent (res. $\mRNA$)), a jump is made to the instruction $I_s$.
Thus in a primitive description of CGF,
every decremental instruction $I_i=\Dec{r_j}{s}$ is described by

\noindent (Decrement instruction $I_i=\Dec{r_j}{s}$)
{
\setlength{\belowdisplayskip}{0pt}%
\setlength{\abovedisplayskip}{0pt}
$$\begin{array}{lccc}
j=1 \hspace{2ex} & 
I_i & =  ! a_1. (0 \bigm | I_{i+1}) \oplus \tau. I_s  & \mbox{with
 $\Dicer=! a_1 .(0 \bigm | I_{i+1} )$} \\
j=2 & 
I_i & =  ! a_2. ( \RISC | I_{i+1}) \oplus \tau. I_s  & \mbox{with
 $\RISC =! a_2 .(\RISC \bigm | I_{i+1} )$}
\end{array}$$
}
The above recursive definition of $\RISC$ for $j=2$
corresponds to the recycling of
	     $\RISC$ described in the degradation.

\smallskip
The decremental instructions so defined contain an error;
that accidental jumps to $I_s$ occur even if the register is non-empty (i.e. in the presence of channel $?a_j$).
This error results from the absence of zero-test of the registers,
a test which cannot be directly formulated
in terms of CGF.
Such an absence has been previously noted by Soloveichik et al.\cite{Soloveichik},
in their studies of stochastic chemical reaction networks.
Lack of zero-test is a main origin of
Turing incompleteness of CGF \cite{ZC},
and motivated Cardelli and Zavattaro to develop their 
Biochemical Ground Form \cite{CZ} as a minimalistic Turing complete
extension of CGF.

\section{Recursive $\RNAI$ and Probabilistic Termination} \label{secrec}
In this section, we model recursive $\RNAI$
in order to improve the defect described in Section \ref{secchem},
that the CGF machine interpretation $\RMI$
allows non-feasible jumps.
We extend the $\RNAI$ mechanism to a recursive $\RNAI$ ($\RNARI$),
whose register machine $\RMRI$ is described in terms of
CGF + fixed points. This interpretation guarantees a probabilistic termination of
the machine.
Via this extended mechanism, $\siRNA$s produced and accumulating during interference
targets not only $\mRNA$
but also $\Dicer$ and $\RISC$.
A schematic of this situation is presented in Figure \ref{reccirc}, in which the usual
$\RNAI$ are displayed to the left, but $\siRNA$s are produced by both $\Dicer$
and $\RISC$ (which simultaneously degrades $\mRNA$). The right hand of Figure
\ref{reccirc} includes
inhibition arrows from $\siRNA$ to
$\Dicer$ and $\RISC$.
The mechanism is recursive because
the $\RISC$ complex containing $\siRNA$
is being degraded besides acting as a degrading agent.
The recursiveness of $\RNAI$
prevents the decrement operators of Section \ref{secchem}
from taking erroneous jumps, since 
$\siRNA$s accumulating throughout the $\RNAI$ cycle work as inhibitors of the decrement
operators.
\begin{figure}[!hbt]
\def\@captype{table}
\begin{minipage}
{0.5\textwidth}
\tiny
\resizebox{8cm}{!}{
$\tiny \xymatrix@!0@C=0.6in@R=1.5pc
{ & &  & \mbox{\sf \small dsRNA} \\
& & 
\ar@{*{|}*{|}*{|}}[rr]  \ar@{-}[rr] <3pt> 
  \ar@{*{|}*{|}*{|}}[rr] <-10pt> \ar@{-}[rr] <-13pt>  & & 
  \ar@{=>}@/^1pc/[ddr]^{\mbox{\normalsize Dicer}}="D"
    & 
&
  \\ \\
\ar@<0pt> [r]^{\mbox{\footnotesize RdRp}}
& \ar@{=>}@/^1pc/[uur]  <2pt>  
 &   &    & &
\ar@{*{|}*{|}*{|}}[l] <13pt> |\hole \ar@{-}[l] <16pt>  |\hole
 \ar@{*{|}*{|}*{|}}[l] <0pt> |\hole \ar@{-}[l] <-3pt>  |\hole 
& 
 \ar@{*{|}*{|}*{|}}[l] <13pt> 
|\hole \ar@{-}[l]<16pt>   |\hole
  \ar@{*{|}*{|}*{|}}[l] <0pt> 
|\hole \ar@{-}[l] <-3pt>_{\mbox{\small
siRNA's}}  |\hole 
&  &  
& 
\ar@{-|}
@/_2pc/
[llluu]
\stackrel{\frac{||||||||}{}}{\mbox{\normalsize $\siRNA$}}  
\ar@{-|}
@/^2pc/
[lllddd]
\\   
   \ar@{*{|}*{|}*{|}}[rr]<9pt> \ar@{-}[rr]<6pt>_{\mbox{\small $\abmRNA$}}  & 
  & & & &  
  \ar@{=>}@/^1pc/[ddl]^{
 \stackrel{\mbox{\footnotesize argonaute}
}{\frac{\mbox{\huge $\frown$}}
{||||||||||||}
}
\mbox{\small \sf RISC}}
   \\  
\\
  &    
  &  \ar@{=>}@/^1pc/[uul]^{
   \mbox{aberration}}  
   &  &
 & 
&
\\  
 & &    
  \ar@{*{|}*{|}*{|}}[rr]<9pt>^{
\frac{
\mbox{\huge $\frown$}
}{|||||||||||||}
}
 \ar@{-}[rr] <6pt>  
  &   \mbox{\sf \small mRNA}  \ar@{=>} @/_1pc/ [dr]
  \ar@{<=} @/^1pc/ [dl]
  &  &  \\
 & &
\mbox{\footnotesize transcription}
  & &  \mbox{\normalsize  $0$, \siRNA's} & & 
}$
}
\caption{Recursive $\RNAI$}
\label{reccirc}
\end{minipage}
\hfill
\begin{minipage}[t]
{0.5\textwidth}
\resizebox{7cm}{!}{
\begin{tabular}
{ll}
(degradation of $\Dicer$) &
\mbox{$\siRNA + \Dicer \longrightarrow 0$} \\
(degradation of $\RISC$) &
\mbox{$\siRNA + \RISC \longrightarrow  0 $}   
\end{tabular}
}
\tblcaption{chemical reactions for $\RNARI$}
\label{recchemreac}
\end{minipage}
\end{figure}

In $\RNARI$, the chemical reactions involved in $\Dicer$
and in $\RISC$ 
are not only those in the second row of Table \ref{chemreac}
but also those in Table \ref{recchemreac}.
The first row of Table \ref{recchemreac} and (cleavage)
represent reciprocal interactions on $\Dicer$ such that $\Dicer$ either
makes $\dsRNA$ disappear by cleavage or $\Dicer$ is degraded by $\siRNA$.
Similar reciprocal interactions for $\RISC$
between the second row of Table \ref{recchemreac} and (degradation).

\smallskip

We next configure $\RNARI$ as
a register machine $\RMRI$ in terms of CGF with fixed points.
\begin{definition}[$\RMRI$ in  CGF with fixed points]\label{RMRICGF}~\\{\em
Registers and $I_i=\Inc{r_j}$ are identical to those of
Section \ref{secchem}. 
The decrement instruction, 
with incorporation of $\siRNA$, is
 {\setlength{\belowdisplayskip}{0pt}%
 \setlength{\abovedisplayskip}{0pt}
 \begin{align*}
 \shortintertext{(Decrement instruction $I_i=\Dec{r_j}{I_s}$) }
 I_i  & =  ! a_j. (0 \bigm | I_{i+1}) \oplus \tau. ( !s. I_{i} \oplus
  \tau. I_s) 
      =   {\sf fix}_{X}. [\,  a. (0 \bigm| I_{i+1}) \oplus \tau.
 (! s.X \oplus \tau. I_s )
    \, ]  \\
 \siRNA & =  ?s. \siRNA \nonumber
 \end{align*}
 }
In the above definition of $I_i$, when $j=1$ (res. $j=2$),
the left term 
$! a_j. (0 \bigm | I_{i+1})$ corresponds to $\Dicer$ (res. $\RISC$)
cleaving $\dsRNA$ (res. degrading $\mRNA$),
while the right term $\tau. ( ! s. I_{i} \oplus I_s)$ corresponds to
$\Dicer$ (res. $\RISC$) being degraded by $\siRNA$.
Hence our definition of $I_i$ intrinsically reflects the
reciprocal interactions of $\Dicer$ and $\RISC$,
and implies a recursive $\RNAI$ process in the presence of $\siRNA$.
}
\end{definition}
The fixed point definition of $I_i$
derives from Zavattaro-Cardelli \cite{ZC},
but here we have highlighted a biological analogue of the definition. In the following, we modify slightly the results of \cite{ZC}
 to obtain the main theorem of this section.

Given a state $(I_i, r_1 = l_1, r_2=l_2)$ of register machine
and a natural number $h$,
the solution in $\RMRI$ is defined by
$
\Mean{(I_i, r_1 = l_1, r_2=l_2)}_h := 
I_i \bigm | \; \prod_{l_1} \dsRNA \; \bigm | \; \prod_{l_2} \mRNA \; \bigm |
\; \prod_h \siRNA
$, 
where $I_i$ on the right hand is that of
Definition \ref{RMRICGF}. 

\begin{prop}[correspondence of computations between machine and
 $\RMRI$] \label{corresp}
Suppose a one step computation of register machine is given by
$
(I_i, r_1 = l_1, r_2=l_2)  \longmapsto (I_j, r_1 = l_1^{'}, r_2=l_2^{'})
$. Then we have the following for the solutions of the two states of the
computation:
\begin{itemize}  \itemsep=0pt \parskip=0pt
\item[-] If $I_i=\Inc{r_j}$ or
$I_i=\Dec{r_j}{s}$ with $l_j=0$,
then the solution $\Mean{(I_i, r_1 = l_1, r_2=l_2)}_h$
can converge to the solution
$\Mean{
(I_j, r_1 = l_1^{'}, r_2=l_2^{'})}^\dag_h$ with
the probability $1$.
\item[-] 
If $l_j> 0$ and $I_i=\Dec{r_j}{s}$,
the solution $\Mean{(I_i, r_1 = l_1, r_2=l_2)}_h$
can reach to a solution
$\Mean{
(I_j, r_1 = l_1^{'}, r_2=l_2^{'})}^\dag_k$
for some natural number $k \geq h+1$ with
the probability $> 1-\frac{1}{h}.$
\end{itemize}
\end{prop}

\noindent{ \bf Proof.} \\
We illustrate the case of $I_i=\Dec{r_j}{s}$
(direct for increment instructions),
where sigma in the second column
denote the probability
that $\RMRI$ computations attain the
right hand side solutions. The schematic in the third column
displays the execution paths for the probability.
$$\begin{array}{|l|cl|l|}
l_j = 0 & \sum_{i=0}^{\infty} (\frac{h}{h+1})^i \times
 \frac{1}{(h+1)} & = 1 &
\xymatrix{
I_i \ar[r]^1  &\ar@<1ex>[l]^h  \bullet \ar[r]^1
& I_s =I_j 
} \\ \hline
l_j \not = 0
&
\sum_{i=0}^{\infty}  (\frac{1}{l_j +1} \times \frac{h}{h+1}
)^i \times \frac{l_j}{l_j +1}  &  >   1- \frac{1}{h}
&
\xymatrix{
I_i \ar[r]^1 \ar[dr]_{l_j} &\ar@<1ex>[l]^h \bullet \ar[r]^1  & I_s 
\\
   &     I_{i+1} = I_j
   &  
      }
\end{array}$$


We now state the main theorem of this section.  
\begin{thm}[probabilistic termination] \label{probterm}
The following are equivalent:
\begin{itemize}  \itemsep=0pt \parskip=0pt
\item[-]
A Minsky register machine starting from a state
$(I_j, r_1 = l_1,  r_2=l_2)$ terminates.

\item[-]
A CGF $(\RMRI,\Mean{(I_j, r_1 = l_1,  r_2=l_2)}_h)$
probabilistically 
terminates with probability
greater than $1-\sum_{k=h}^\infty \frac{1}{k}$. 
\end{itemize}
\end{thm}

\noindent{ \bf Proof.}
Note first that following the execution of a decrement instruction,
the number of $\siRNA$ increases by at lease one.
This is because at least one
$\siRNA$ is produced by $\Dicer$ cleavage or by $\RISC$
(as it degrades $\mRNA$). By Proposition \ref{corresp}
a computation of register machine containing $d$ decrement
instructions is faithfully reproduced with probability greater than
the following:
$(1-\frac{1}{h})(1-\frac{1}{h+k_1}) \cdots (1-\frac{1}{h+k_1+\cdots+ k_d})
\geq \prod_{k=h}^{h+d } (1-\frac{1}{k})  > 
1 - \sum_{k=h}^{h+d } \frac{1}{k}
$
, where $k_i \geq 1$ is the number of $\siRNA$s produced
by the corresponding decrement instruction.
\hfill 
$\Box$


\bibliographystyle{eptcs}


\end{document}